\documentclass[apj, iop]{emulateapj}

\usepackage{graphics,graphicx}
\usepackage{natbib}
\catcode`\@=11
\def\gsim{\ifmmode{\mathrel{\mathpalette\@versim>}}
    \else{$\mathrel{\mathpalette\@versim>}$}\fi}
\def\lsim{\ifmmode{\mathrel{\mathpalette\@versim<}}
    \else{$\mathrel{\mathpalette\@versim<}$}\fi}
\def\@versim#1#2{\lower 2.9truept \vbox{\baselineskip 0pt \lineskip
    0.5truept \ialign{$\m@th#1\hfil##\hfil$\crcr#2\crcr\sim\crcr}}}
\catcode`\@=12

\def\mstar{M_\ast}
\def\mgas{M_g}
\def\mbh{M_{BH}}

\def\sigv{\sigma_v}
\def\sigvsq{\sigv^2}
\def\rv{r_v}

\def\mstarone{M_{\ast 1}}
\def\mgasone{M_{g1}}

\def\sigvone{\sigma_{v1}}
\def\sigvonesq{\sigma_{v1}^2}

\def\mstartwo{M_{\ast 2}}
\def\mgastwo{M_{g2}}

\def\sigvtwo{\sigma_{v2}}
\def\sigvtwosq{\sigma_{v2}^2}

\newcommand{\msun}{\ensuremath{{{\rm M}}_{\scriptscriptstyle \odot}}}
\newcommand{\kms}{\,{\rm km\,s^{-1}}}

\newcommand{\beq}{
\begin{equation}
}
\newcommand{\eeq}{
\end{equation}
}

\def\re{R_{\rm e}}

\shorttitle{Black holes in  Central Cluster Galaxies}
\shortauthors{Volonteri \& Ciotti}

\begin{document}

\title{Massive black holes in Central Cluster Galaxies}

\author{Marta Volonteri\altaffilmark{1,2} \& Luca Ciotti\altaffilmark{3}}

\altaffiltext{1}{Institut d'Astrophysique de Paris, 98bis Bd. Arago, 75014 Paris, France}
\altaffiltext{2}{Department of Astronomy, University of Michigan, Ann Arbor, MI, USA}
\altaffiltext{3}{Dipartimento di Fisica e Astronomia, Universit\`a di Bologna,  via Ranzani 1, 40127 Bologna, Italy}

\begin{abstract}
We explore how the co-evolution of massive black holes (MBHs) and galaxies is affected by environmental effects, addressing in particular MBHs hosted in the central cluster galaxies  (we will refer to these galaxies in general as `CCGs'). Recently the sample of MBHs in CCGs with dynamically measured masses has increased, and it has been suggested that these MBH masses  ($M_{\rm BH}$) deviate from the expected correlations with velocity dispersion ($\sigma$) and mass of the bulge ($M_{\rm bulge}$) of the host galaxy: MBHs in CCGs appear to be `over-massive'. This discrepancy is more pronounced when considering the $M_{\rm BH}-\sigma$ relation than the $M_{\rm BH}-M_{\rm bulge}$ one. We show that this behavior stems from a combination of two natural factors, (i) that CCGs  experience more mergers involving spheroidal galaxies and their MBHs, and (ii) that such mergers are preferentially gas-poor.  We use a combination of analytical and semi-analytical models to investigate the MBH-galaxy co-evolution in different environments and find that the combination of these two factors is in accordance with the trends observed in current data-sets.

\end{abstract}


\section{Introduction} 
The discovery of correlations between MBHs and their hosts \citep{Magorrian1998,fm00,Gebhardt2000} has been taken as one of the main elements in support of a co-evolution between MBHs and galaxies, that in turn brought to the suggestion that energy input from an accreting MBHs, as quasar or Active Galactic Nucleus (AGN) may regulate star formation in the host,  or create a symbiosis between MBH and stellar growth. Indeed, theoretically there is reason to expect that energy or momentum driven outflows establish correlations that scale as  $M_{\rm BH} \propto \sigma^5$ and $M_{\rm BH} \propto \sigma^4$ respectively \citep{Silk1998,Fabian1999}. The correlations between MBHs and galaxies, derived on a sample of about 50-70 MBHs \citep[e.g.,][]{Gultekin2009,2011MNRAS.412.2211G,McConnell2012a} are also extrapolated to the whole population, assuming that each galaxy hosts a MBH consistent with the correlation and its scatter to derive global properties, such as the mass density in MBHs today \citep[e.g.,][]{Shankar2004}. These correlations, or deviations from such correlations, also appear to hold the key to understanding the formation and evolution of the MBH population \citep[and references therein]{Volonteri2012Sci}. 

Recent measurements of the masses of MBHs in CCGs (we include in this definition central dominant galaxies as well as brightest cluster galaxies) seem to suggest that these MBHs are `over-massive' compared to expectations from of the $\mbh-\sigma$ relation, but they appear more consistent, albeit with a large scatter, with the $\mbh-L_V$ relation \citep{McConnell2011,McConnell2012}. \cite{McConnell2012a} and \cite{2012arXiv1211.3199G} fit  for the  $\mbh-\sigma$ correlation finding a steep correlation, e.g, from  $M_{\rm BH} \propto \sigma^{4.2}$ \citep{Gultekin2009} to $M_{\rm BH} \propto \sigma^{5.5-5.6}$. The same does not happen for the  $M_{\rm BH}-M_{\rm bulge}$ relation, that appears to be consistent with being linear in all cases (although including MBHS in CCGs increases the normalization).    
There is also tentative evidence for over-massive MBHs in CCGs from comparisons with the expectations from the  Fundamental Plane of black hole activity \citep{Hlavacek2012}.  Feedback affecting galaxies in different ways depending on potential well \citep{Booth2011,2005MNRAS.358..168S} and environment, via gas-rich mergers and cooling flows \citep{Zubovas2012} has been proposed to be partially responsible for the properties of these MBHs. 

We investigate here two guiding ideas related to the influence of mergers between spheroidal, gas-poor galaxies  (`dry' mergers) and of MBH-MBH mergers to determine the astrophysical drivers of the possibly different relationship between MBHs and their hosts in the case of CCGs.  It is well established from the theoretical point of view that (parabolic) dry mergers consistently grow a galaxy's mass, luminosity and radius more than a galaxy's velocity dispersion \citep[e.g.,][] {CvA2001,Nipoti2003,CLV2007,Naab2009,Nipoti2009,Shankar2011,Oser2012,Hilz2012}, especially when a galaxy is dominant over the general population  \citep{CLV2007,Naab2009}. Qualitatively this implies that eventually the largest galaxies deviate from the $\mbh-\sigma$ relation. Therefore, if CCGs grew predominantly through  dry mergers  \citep[e.g.,][]{Ostriker1975,Hausman}, and their MBHs grew mostly through MBH-MBH mergers  \citep{Malbon2007,Yoo2007} in such dry mergers then the expectation is that they will be outliers  in the $\mbh-\sigma$ relation \citep[see also][]{Boylan2006,Lauer2007,Zhang2012}. Support for the influence of dry mergers in shaping the structural properties of CCGs comes from their steeper radius--luminosity relation and flatter Faber-Jackson relation \citep[e.g.,][]{Lauer2007,Bernardi2007a,Desroches2007}. Additionally, if MBHs in CCGs experience a significant mass increase because of MBH-MBH mergers with respect to MBHs in a field environment, the extra boost in mass may also contribute to explain the properties of MBHs in CCGs, if gas accretion is responsible for establishing the relationships in the first place. 

We present here models of MBH growth in galaxies that take into account both the cosmic environment (the frequency and properties of galaxy and MBH mergers) as well as the evolution of the structural properties of galaxies during galaxy mergers, with particular attention to mergers between gas-poor spheroids.  


%
%

\section{Black hole growth}
We use the latest incarnation of our semi-analytical models \citep{Volonteri2012} to provide a simple estimate of the influence of dry mergers on the build-up of MBHs and their hosts in different environments.  We start from Monte Carlo realizations of the  merger hierarchy of dark matter halos from early times to the present in a $\Lambda$CDM cosmology (WMAP5). The mass resolution is set to be $10^{-3}$ times the mass of the main halo at $z=0$, and it reaches $10^5\,\msun$ at $z=20$.   

We summarize here the main assumptions of the semi-analytical model.  We wish to keep our models as simple as possible, while making sure that the properties of the MBHs we study are correctly determined through the cosmic evolution of their hosts. We do not explicitly model in detail the complex physics of the gaseous component of galaxies through cooling, star formation and various feedbacks \citep[see][and references therein for models that treat in detail semi-analytically the baryonic component of galaxies and its link to MBH evolution]{Fanidakis11,Fontanot2011,Hirschmann2012,2012MNRAS.423.2533B}  even though a simple scheme taking into account gas dissipation and star formation is implemented as described in Section 3. In particular, in the present approach we follow the evolution of the three relevant cosmological components of interest, namely 1) the dark matter halos, 2) the galaxies (stars plus gas), and 3) the central MBHs. We briefly describe how their properties are determined as a function of the merging events.

In practice, each  halo is modeled as an isothermal sphere of one-dimensional velocity dispersion $\sigma_{\rm vir}$ and circular velocity $v_{circ}=\sqrt[]{2}\sigma_{\rm vir}$. Each halo is also characterized by a virial radius $r_{h}$ and a mass $M_h$ contained within $r_{h}$, such that the mean density within the virial radius is $\Delta_{\rm vir}\,\rho_{\rm crit}$, where  $\rho_{\rm crit}$ is the critical density for closure at the redshift of collapse and $\Delta_{\rm vir}$ is the density contrast at virialization for the chosen cosmology. When two halos merge, the new total mass inside the virial radius is computed as the sum of the two participating masses, and the new virial radius is determined by requiring that the mean halo density within the new $r_h$ is $\Delta_{\rm vir}$ times the crititical density at that redshift. Therefore, the new $v_{circ}$ and $\sigma_{\rm vir}$ are uniquely determined. As described in the following, $\sigma_{\rm vir}$ is a fundamental ingredient in order to determine accretion on MBHs.

Galaxy morphology is related  to the merger history, using a three-parameter model, where spheroid (we equivalently refer to spheroids and `gas-poor' galaxies in the following) formation depends on both the mass ratio of the two merging halos and their velocity dispersion, and the timescale over which a spheroid can re-acquire a disc through cold flows and mergers with gas-rich galaxies.   \cite{Koda2007} show that the fraction of disc- vs spheroid-dominated galaxies is well explained if the only merger events that lead to spheroid formation have mass ratio $>$0.3 and  $\sigma_{\rm vir}>55 \kms$; also, the merger timescale \citep[calculated following][]{Boylan2008} must be less than the time between the beginning of the merger and today, $z=0$. We assume that spheroids form after a merger that meets these requirement.  In order to include the effect of gas accretion and cold flows we additionally allow gas to re-condense (`gas-rich' galaxy) after 5 Gyrs in galaxies with $\sigma_{\rm vir}<300 \kms$ where no major mergers occurred. We not explicitly add an amount of re-accreted gas, we simply label the galaxy as ``gas-rich".  In Section 3 we describe in detail how ``bona fide" structural galaxy properties are determined following a merger and the associated gas dissipation.

At high redshift we seed dark matter halos with MBHs created by gas collapse whose mass is linked to the halo $v_{circ}$. Specifically, we adopt here the formation model detailed in Natarajan \& Volonteri (2011) based on Toomre instabilities (Lodato \& Natarajan 2006).  The subsequent evolution of MBHs   includes MBH-MBH mergers, merger-driven gas accretion, stochastic fueling of MBHs through molecular cloud capture,  a basic implementation of accretion of recycled gas from stellar evolution.  

We  assume that, when two galaxies hosting MBHs merge, the MBHs themselves merge within the merger timescale of the host halos \citep[and references therein]{Dotti2007}. We also include gravitational recoils using the formalism described in \cite{campanellietal07}. Note that in our models MBHs accrete mass whenever it is available to them. At the time of a MBH-MBH merger we just sum the masses at that specific time. They may accrete mass driven by the galaxy merger or through accretion of gas cloud before the MBH-MBH merger proper, during (i.e., at the same timestep) or after, i.e. on the merged MBH (as long as the MBH is not ejected from the galaxy center). Since our timesteps are typically short compared to the Salpeter time ($\sim 10^5-10^6$ yrs) we can keep track of accretion (including the evolution of the accretion rate) and mergers self-consistently rather than simply adding all the accreted mass at one single timestep. 

After a halo merger with mass ratio larger than 3:10, in which at least one of the two halos is the host of a gas-rich galaxy, we assume that a merger-driven accretion episode is triggered. The accretion rate is set at the Eddington limit until the MBH mass reaches $M_{\rm BH,\sigma} \, =\, 5\times10^7  \left({\sigma_{\rm vir}}/{200 \kms} \right)^4 \, \msun$. We therefore assume that it is the potential well of the host halo that sets the maximum mass at which the MBH can grow through high accretion rate accretion.  After that, self-regulation ensues and the MBH feedback unbinds the gas closest to the MBH, thus reducing its feeding.  We model the decrease of the accretion rate as:
\beq
f_{\rm Edd}(t)=\left(\frac{t+t_{f_{\rm Edd}}}{t_{f_{\rm Edd}}}\right)^{-\eta_L},
\label{decay}
\eeq
where $\eta_L \simeq2$ and $t_{f_{\rm Edd}}\simeq4.1\times 10^6(M_{\rm BH,\sigma}/10^8\msun)$ yr, and $t=0$ (where $f_{\rm Edd}=1$) represents the time when the MBH reaches the threshold for self-regulation \citep{HopkinsHernquist2006}.  Since the accretion rate $\dot{M}_{\rm in}(t)=f_{\rm Edd}(t)\,M_{\rm BH}(t)/t_{\rm Edd}$ is time-dependent, it is  integrated self-consistently at each timestep to obtain the growth of the MBH mass until the accretion rate drops below $f_{\rm Edd}\simeq 10^{-5}$, at which point we consider the accretion episode concluded. 
 
In gas-rich galaxies we include also feeding through molecular clouds (MCs). At each timestep, $\Delta t$, we determine the probability of an accretion event as:  
\beq
{\cal P}=\frac{\Delta t}{t_{MC}}\simeq \frac{10^{-3} \sigma_{\rm vir} {\Delta t}}{R_{\rm cl}}
\label{prob}
\eeq
where $R_{\rm cl}\simeq 10$ pc  (Hopkins \& Hernquist 2006).  We assume a lognormal distribution for the mass function of clouds close to galaxy centers \citep[peaked at $\log(M_{\rm MC}/$\msun$)=4$, with a dispersion of 0.75 based on the Milky Way case, e.g.,][]{Perets2007}.   In order to derive the accretion rate and duration of the accretion episode caused by the MC, we derive the properties of the accretion disc created in one of these events (including disc size, $R_d$, mass, $M_d$, accretion rate and duration of the accretion episode). We  assume that the MBH captures only material passing within the Bondi radius, $R_B$, and that its specific angular momentum is conserved (Bottema \& Sanders 1986; Wardle \& Yusef-Zadeh 2008).
The outer edge of the disc that forms around the MBH corresponds to the material originally at $R_B$:
\beq
R_d=2 \lambda^2 R_B=8.9\, {\rm pc}\,\lambda^2 \frac{M_{BH}}{10^7\msun}\left(\frac{\sigma_{\rm vir}}{100 \kms}\right)^{-2}, 
\eeq
where $\lambda$ is the fraction of angular momentum retained by gas during circularization. The maximum captured mass will be contained in a cylinder with radius $R_B$ and length $2\times R_{\rm cl}$, the MC diameter. If $\kappa$ is the ratio of the mass going into the disc with respect to  the whole mass in the cylinder, then:
\begin{eqnarray}\nonumber
M_{\rm d}&=& \kappa \frac{R_B}{R_{\rm cl}}M_{\rm MC}= 4.7\times 10^4  \kappa  \left(\frac{M_{BH}}{10^7\msun}\right)^2\\
&& \left(\frac{\sigma_{\rm vir}}{100 \kms}\right)^{-4} \msun. 
\label{MdiscMC}
\end{eqnarray}

We assume that the disc is consumed over its viscous timescale:
\begin{eqnarray}\nonumber
t_{\rm visc}&=&\left(\frac{R^{3}_d}{\alpha_v^2 GM_d}\right)^{1/2}=3.4 \times 10^5 \frac{\lambda^3}{\alpha_v}\frac{M_{BH}}{10^7\msun}\\
&& \left(\frac{\sigma_{\rm vir}}{100 \kms}\right)^{-3}  {\rm yr},
\end{eqnarray}
so that we can calculate an upper limit to the mean accretion rate:
\begin{eqnarray}\nonumber
\dot{M}_{\rm max}&=&\frac{M_{\rm d}}{t_{\rm visc}}=0.13 \frac{\alpha_v \kappa}{\lambda^3}\frac{M_{BH}}{10^7\msun}\\
&& \left(\frac{\sigma_{\rm vir}}{100 \kms}\right)^{-1} N_{23} \,\msun {\rm yr}^{-1} , 
\end{eqnarray}
where $N_{23}$ is the column density in the cloud in units of $10^{23}$ cm$^{-2}$, and $\alpha_v=0.1$ is the viscosity parameter.  Following Wardle \& Yusef-Zadeh (2012) we  set $\lambda=0.3$ and $\kappa=1$. If $\dot{M}_{\rm max}$ is less than the Eddington rate (assuming a radiative efficiency of 10\%) we let the MBH accrete the whole $M_{\rm d}$ over a time $t_{\rm visc}$, otherwise we treat accretion similarly to the ``decline" phase of quasars, as feedback from the high luminosity produced by accreting the cloud will limit the amount of material the MBH can effectively swallow. 

In gas-poor galaxies,  feeding of a MBH can be sustained by the recycled gas (primarily from red giant winds and planetary nebulae) of the evolving stellar population \citep[e.g.,][]{Ciotti1997,2012MNRAS.427.2734N}. The geometrical model by \cite{Volonteri2011} implies that the quiescent level of accretion onto a central MBH due to recycled gas is $f_{\rm Edd}\simeq 10^{-5}$.

%
%
%
%

\begin{deluxetable}{cccc}
\tablenum{1}
\tablehead{\colhead{$M_{\rm halo}$ } & \colhead{Dry mergers} & \colhead{Dry mergers } & \colhead{$M_{\rm BH,0}/M_{\rm acc}$} \\ 
\colhead{($M_\odot$)} & \colhead{ with MBHs} & \colhead{(total)} & \colhead{}  } 

\startdata
$10^{15}$ & 3.3 ($\sigma_n$=1.8) & 3.5 ($\sigma_n$=2.1)  &	 10.9 $\pm$ 9.7 \\
$10^{14}$ & 1.5 ($\sigma_n$=1.2) & 1.6 ($\sigma_n$=1.2) & 	 2.4 $\pm$ 1.0\\
$4\times10^{13}$ & 0.4 ($\sigma_n$=0.6) & 0.4 ($\sigma_n$=0.6) & 1.9 $\pm$ 0.7 \\
$2\times10^{13}$ & 0.1 ($\sigma_n$=0.4) & 0.2 ($\sigma_n$=0.4) &2.0  $\pm$ 0.7\\
$10^{13}$ & 0.1 ($\sigma_n$=0.3) & 0.1 ($\sigma_n$=0.3)& 	1.8 $\pm$ 1.8 \\
 &  \\
\enddata
\tablecomments{Number of  mergers (including their $1-\sigma$ variance) with mass ratio $>0.1$ experienced by the central galaxy, regardless of its morphology, of a halo that by $z=0$ has the mass listed in Column 1, and growth channels for its MBH.}
\end{deluxetable}

Through this model we can disentangle what the main contributions to the growth of MBHs and galaxies in different halos is.  As regards galaxies, in Table~1 we list the number of dry mergers experienced over its lifetime by the central galaxy of a halo, as a function  of the halo mass at $z=0$ (the average is performed over 20 to 35 different halos for each halo mass). We can consider the  $10^{15}\, \msun$ halo as a cluster-sized halo, the $10^{14}\, \msun$ halo as a group-sized halo, and the smaller halos as galaxy-sized ones.  The dry-ness of a galaxy during a merger is determined in a very simple way, i.e. we require that both merging galaxies are gas-poor.  As a reference the total number of mergers (dry and wet) with mass ratio $>0.1$ experienced by the same galaxies since $z=4$ is typically between 8 and 12. As expected, the number of dry mergers decreases as the halo mass decreases, therefore their influence will be strongest for central galaxies in clusters, and possibly in groups. As highlighted in Column 2, most, but not all, dry mergers involve galaxies both hosting a MBH. 

\begin{figure}
\includegraphics[width= \columnwidth]{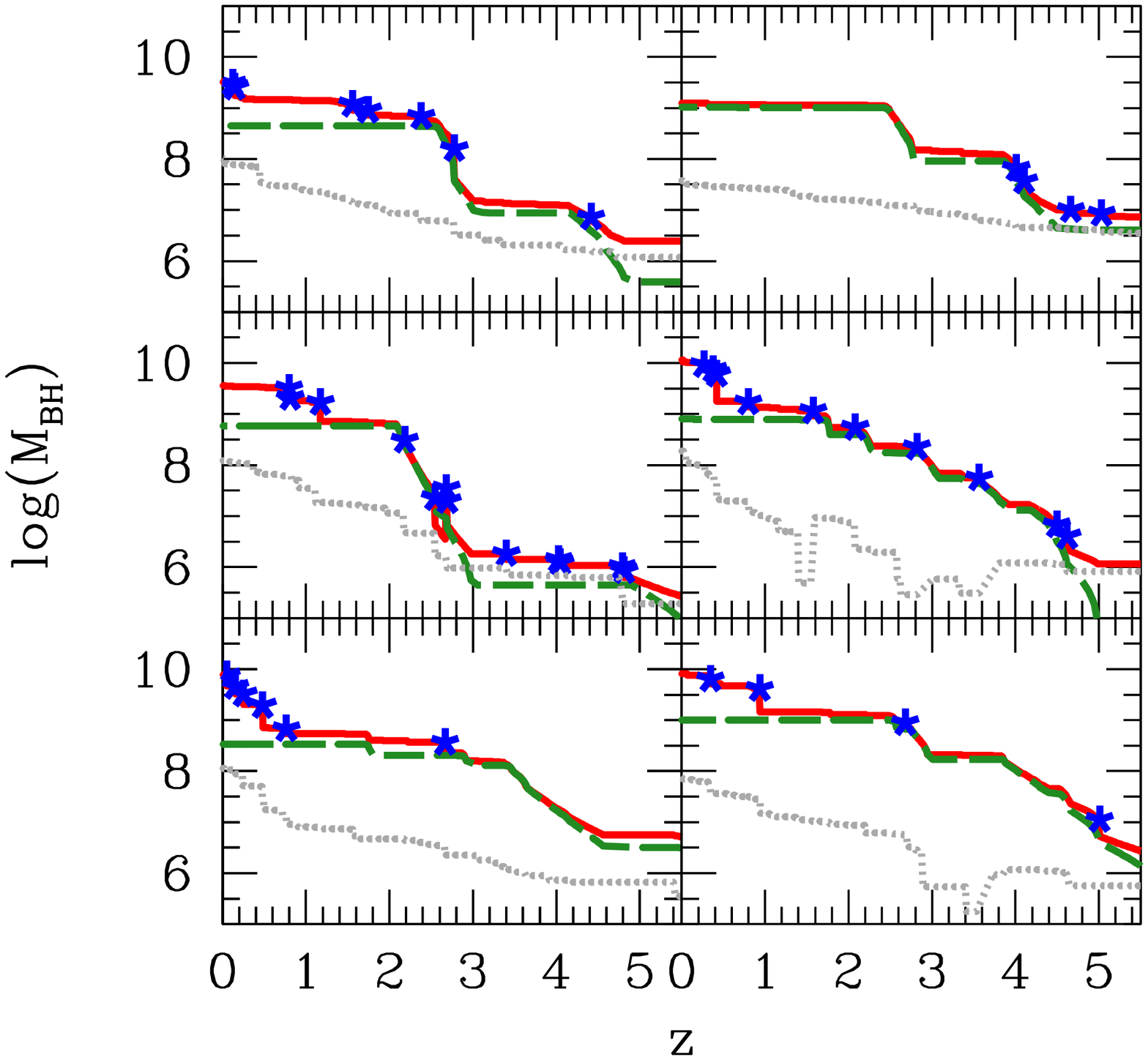}
\caption{Examples of MBH growth in six randomly chosen central galaxies of $10^{15} \,\msun$ halos.  The red solid curve shows the MBH mass at a given time, the dark green dashed curve shows the cumulative mass gained in accretion events, blue asterisks mark MBH-MBH mergers with mass ratio $>1:10$.  The dotted grey curve shows the curves we obtained for the same MBHs when we ``turned-off'' all accretion channels, so that the only growth mechanism is through MBH-MBH mergers (the decrease in MBH masses in such curves are due to dynamical ejections of MBH, not to evaporation or mass loss). As an example, the growth of the MBH in the top-left panel is dominated by MBH-MBH mergers, while that of the MBH in the top-right panel is dominated by accretion through various channels (merger driven in gas-rich mergers, through accretion of gas clouds, and through recycled gas).
\label{fig:BHgrowth_BCGs}}
\end{figure}

In Fig.~\ref{fig:BHgrowth_BCGs} we focus on the MBH growth in six different realizations of $10^{15}\, \msun$ halos.  For instance, the growth of the two MBHs in the top panels is strikingly different. In one case it is dominated by MBH-MBH mergers at $z<2$, while the second MBH does not experience any important  MBH-MBH merger at late cosmic time.  If we define the relative growth through mergers as the difference between the MBH mass at $z=0$, $M_{\rm BH,0}$, and the cumulative accreted gas mass, $M_{\rm acc}$, i.e. we sum all the gas mass that is added to this MBH in all accretion episodes throughout its lifetime, we can estimate the importance of MBH-MBH mergers in a MBH's history.  In Table ~1 we report this information in the fourth column. Note that the definition we adopt does {\it not} account for gas accretion on the MBHs that merge with the one in the central galaxy. Therefore, the fact that in general $M_{\rm BH,0}/M_{\rm acc}>1$ does {\it not} mean that MBHs grow predominantly through mergers, and in fact if we ignored all accretion onto MBHs from our models, the final MBH mass would be about two orders of magnitude smaller (cf. gray dotted curves in Fig.~1). Overall, therefore, MBHs grow through accretion of gas, in a way consistent with independent  estimates \citep{YuTremaine2002}.

To test the robustness of our results we changed the MBH feeding scheme, by assuming that all MBH accretion activity is driven by galaxy mergers, that the accretion rate is fixed to 30\% of the Eddington rate and that accretion stops once the MBHs have reached the value of $M_{\rm BH,\sigma}$. We find that while small quantitative differences exist, qualitatively the results of our investigation are unchanged. We conclude that, if the $\mbh-\sigma$ relation is established because of accretion-driven feedback \citep[e.g.,][]{Silk1998,Fabian1999}, MBHs in CCGs may not necessarily obey the same the $\mbh-\sigma$ relation, as their masses may grow substantially through mergers after accretion dwindles.

\section{Mergers and galaxy structure}
From the evolutionary histories of our models we extract the series of mergers that the central galaxy of the main halo experiences from $z=1$. We record the MBH mass in the merging galaxies and the dark matter halo masses within the virial radius, $M_{h1}$ and $M_{h2}$, (here onwards the subscripts refer to each one of the two galaxies, with the convention that subscript $1$ labels the central galaxy of the main halo, which is not necessarily the most massive galaxy of the merging pair). To each halo at the starting redshift we assign a stellar mass, $\mstar$, using the fits by Behroozi et al. (2010, see also Nipoti et al. 2012).  If a galaxy is identified as gas-rich we assume that only 75\% of the stars are distributed in the spheroidal part. We then assign a mass-to-light ratio,  a projected velocity dispersion, $\sigma$, consistent with the Faber-Jackson relation, and an effective radius determined through the Fundamental Plane \citep[pages 23-24]{binney1987}, including scatter in all relations. We included a redshift dependence of the Faber-Jackson and Kormendy relations, using the scalings suggested by \cite{Oser2012}, namely that  $\sigma\propto(1+z)^{0.44}$ and $\re\propto(1+z)^{-1.44}$ (see also Nipoti et al. 2012).  We then calculate  $\mstar$, $R_e$ and $\sigma$ resulting from the merger as follows (Ciotti et al. 2007).   We account for weak homology by relating $\sigma$ to  the galaxy virial velocity dispersion, $\sigma_v$, and $R_e$ to the galaxy virial radius, $r_v$ , as:
\begin{equation}
{\sigma\over\sigv}\simeq {24.31+1.91 n +n^2\over 44.23+0.025 n +0.99 n^2};
\end{equation}
\begin{equation}
{\rv\over\re}\simeq {250.26+7.15 n\over 77.73+n^2},
\end{equation}
adequate for Sersic index $n$, $2\lsim n \lsim 12$.  We assume that  the Sersic index of the resulting galaxy is $n=1+\max(n_1,n_2)$, where $n_1$ and $n_2$ are the Sersic indices of the progenitors (each galaxy starts with $n$ determined  inverting the virial identity $G\mstar/\re\sigma^2=(\rv/\re)\times (\sigv/\sigma)^2$ for given $\sigma$ and $\re$). 

In case of  mergers between gas-rich galaxies or between a gas-poor and a gas-rich galaxy, we assume that each gas-rich galaxy has a gas mass $\mgas =\alpha \mstar$, with $\alpha=4$ for all galaxies independently of their histories and that a fraction $\eta=0.05\,(M_{h2}/M_{h1})$ of the gas is converted into stars: 
\begin{equation}
\mstar = \mstarone +\mstartwo +\eta (\mgasone +\mgastwo).
\label{eq:mstar}
\end{equation}
We find the velocity dispersion of the newly formed galaxy\footnote{The presence of  dark matter can be included through an extra $\alpha_{DM}$ parameter in our scheme (see Ciotti et al. 2007), assuming that stars and dark matter are similarly distributed. Inclusion of the dark matter halo represents a small correction as long as we are interested in the region within the effective radius.} as:
\begin{equation}
\sigvsq = {M_{\rm gal1}\over M_{\rm gal}}A_1\sigvonesq +
          {M_{\rm gal2} \over M_{\rm gal}}A_2\sigvtwosq,
\label{eq:sigp}
\end{equation}
where $M_{\rm gal1}=\mstarone+\mgasone$, $M_{\rm gal2}=\mstartwo+\mgastwo$, $M_{\rm gal}=M_{\rm gal1}+M_{\rm gal2}$, and 
\begin{equation}
A_1 = 1 + {\eta\alpha_1\over 1+\alpha_1},
\label{eq:Ai}
\end{equation}
and a similar expression holds for $A_2$.  From the second merger of the sequence onwards, the central galaxy retains the properties derived in the previous step, while we assign $\mstartwo$, and $\sigvtwo$ to the merging galaxy as described above. Throughout our experiment, the MBH mass is determined from the semi-analytical model, that includes both accretion and MBH-MBH mergers. We note that not all galaxies host MBHs, i.e., the merging galaxy may or may not contribute to the MBH growth.

\begin{figure}
\includegraphics[width= \columnwidth]{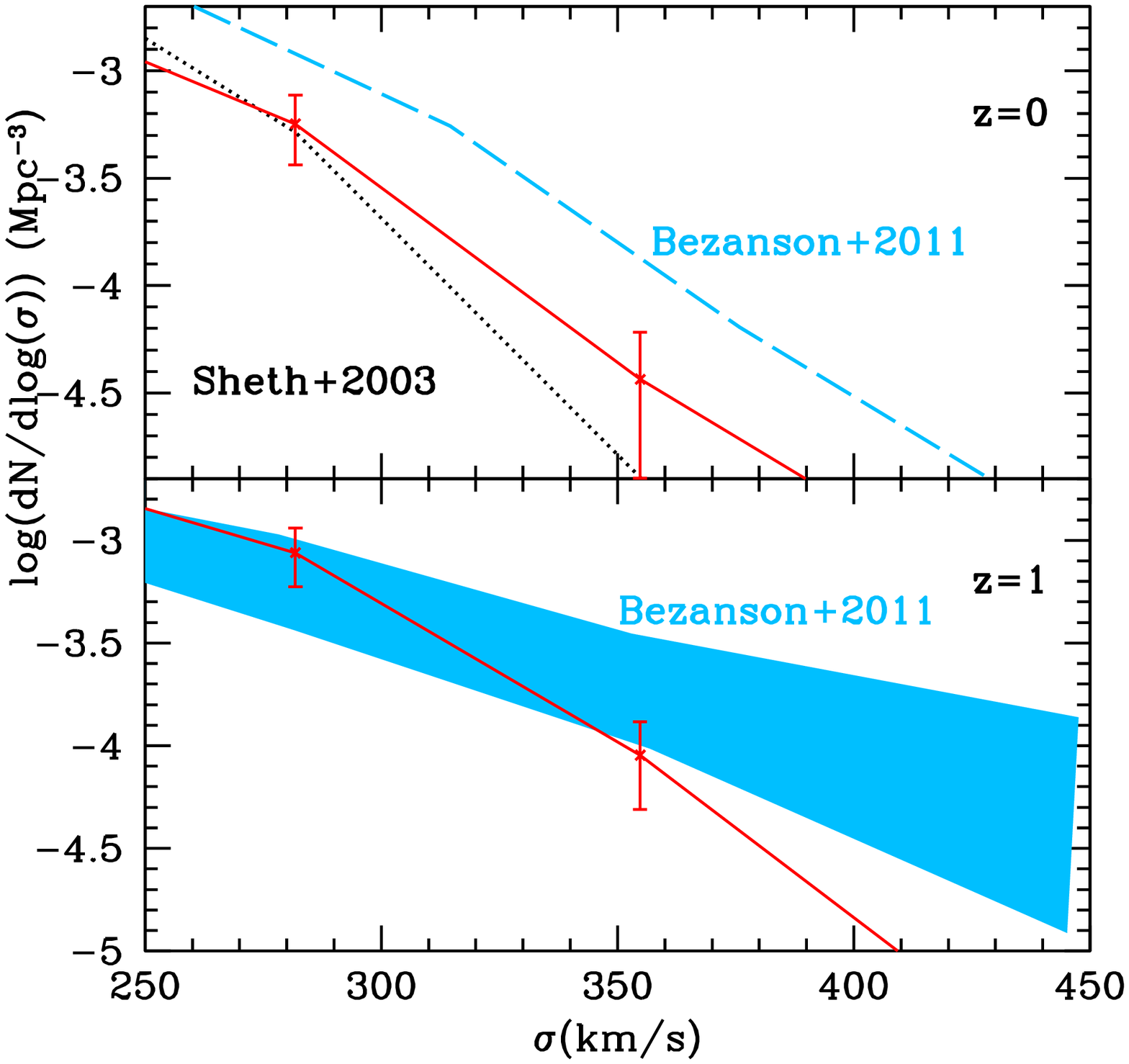}
\caption{Velocity dispersion distribution functions at $z=0$ (top), and $z=1$ (bottom). The model (red solid curve) is compared to distribution functions derived from different surveys  (at $z=0$ we show the function derived by Sheth et al. 2003, as well as the correction introduced by  Bezanson et al. 2011 to account for scatter. At $z=1$ we report the range of values from Fig.~3 in Bezanson et al. 2011).}
\label{fig:VDF}
\end{figure}

\begin{figure}
\includegraphics[width= \columnwidth]{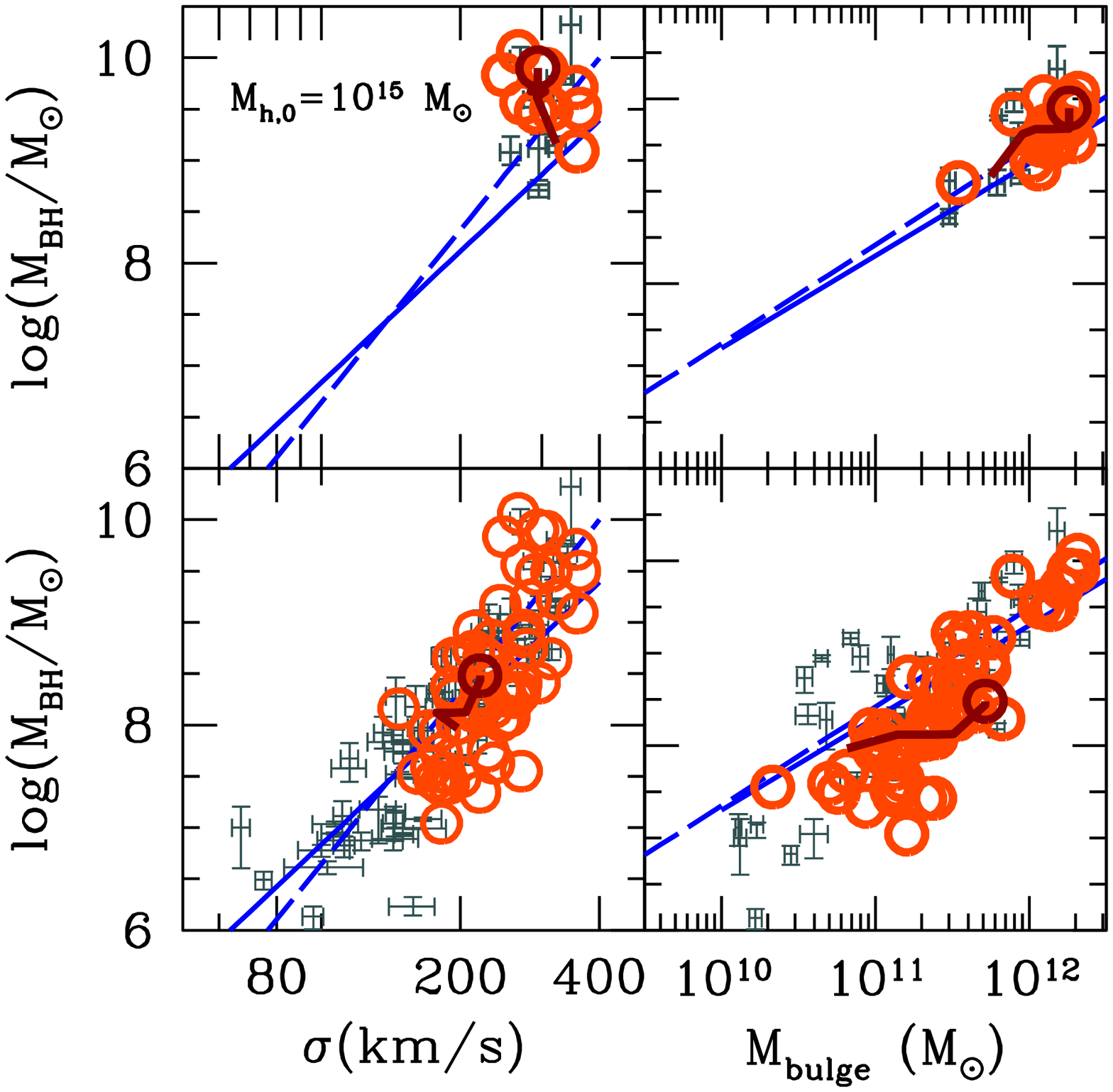}
\caption{Examples of $\mbh-\sigma$ and $\mbh-M_{\rm bulge}$  in 20 randomly chosen central galaxies of $10^{15}\, \msun$ (top) and $10^{13}--10^{15}\, \msun$ (bottom) halos. The gray errorbars are observed MBHs and galaxies \citep[in the top panel we show CCGs only]{McConnell2012a}, the solid blue lines are the fits derived by \cite{Gultekin2009} and \cite{MarconiHunt2003}, the dashed lines are the fits from \cite{McConnell2012a}.  The red dots  show the location of model MBHs at $z=0$. The dark red curves show two examples from the evolutionary histories:  horizontal rightward swings in the $\mbh-\sigma$ panels occur when the galaxy merges with a more massive galaxy, vertical  (leftward) movement is characteristic of dry mergers with similar (smaller) galaxies.}
\label{fig:seq}
\end{figure}

 We present the results of this experiment in Fig.~\ref{fig:VDF} and~\ref{fig:seq}. In  Fig.~\ref{fig:VDF}  we compare the velocity dispersion distribution function of modeled galaxies to distribution functions at $z=0$ and $z=1$ (Sheth et al. 2003; Bezanson et al. 2012), showing an adequate agreement, except for an underestimate of galaxies with very high $\sigma$ at $z=1$.  In the top two panels of Fig.~\ref{fig:seq} we focus on MBHs hosted in the central galaxy of $10^{15}\, \msun$ halos at $z=0$ (akin to CCGs), while in the bottom panel we include all MBHs in halos  with mass $>10^{13}\, \msun$ at $z=0$. At lower masses secular processes are likely to dominate over mergers and our model is less suitable to describe their evolution. We assessed that the results at $z=0$ are qualitatively unchanged if we ignore the redshift evolution in the Faber-Jackson and Kormendy relations, with MBHs and galaxies occupying the same region in the $\mbh-\sigma$ relation. CCGs are characterized by MBHs that consistently deviate from the expected correlations, being over-massive at fixed galaxy properties, occupying the same range as observations. MBHs in elliptical galaxies that are not CCGs, where the influence of MBH-MBH mergers and dry mergers is much milder, tend instead to sit closer to the global MBH-host correlations.  This result can be interpreted as  overall steeper and higher normalized relations with respect to previous estimates (compare solid and dashed black lines in Fig.~2).  At $z<1$ most CCGs are the dominant galaxies, and they merge with smaller galaxies that consistently have $\sigvtwo<\sigvone$ and therefore their velocity dispersion cannot increase (see Eq.~\ref{eq:sigp}). On the other hand a non-CCG galaxy has a higher chance of merging with a galaxy with $\sigvtwo>\sigvone$, with the merger remnant having $\sigv>\sigvone$ (see the dark red tracks in Fig.~2 for an example).

We note \citep[see also][]{Nipoti2012} that this scheme tends to overproduce stellar masses by $z=0$. In fact the  $M_{\rm star}-M_{\rm h}$ relationship peaks at $M_{\rm h}=10^{12}\, \msun$, therefore by merging galaxies close to the peak the remnant galaxy ends up having an increased $M_{\rm star}-M_{\rm h}$. While our merger sequence includes both galaxies below, at and above the peak, we find that we consistently overpredict stellar masses at $z=0$ with respect to the scaling we would obtain directly from the  $M_{\rm star}-M_{\rm h}$ relationship, and that in general bulges appear too massive (this is evident in the bottom-right panel of Fig.~\ref{fig:seq}). This problem would be alleviated if we included corrections for mass lost in the merging process \citep{Nipoti2003}. 
We also consider parabolic mergers only, where  energy is perfectly conserved. However, not all mergers involving a CCG are necessarily parabolic. On the one hand, since the galaxies move in the potential of the cluster, hyperbolic mergers may occur. On the other hand,  \cite{Nipoti2003b} note the effect of dynamical friction, that braking the galaxy's orbit may induce elliptical merging. Hyperbolic and negative-energy mergers  involving a bound pair have competitive effects on the evolution of $\sigma$.  Mergers with negative orbital energy increase the final $\sigma$, and viceversa.  \cite{Hilz2012}  also suggest that  energy transfer from bulge to halo grows the velocity dispersion further.

\section{Conclusions}
In this paper we have highlighted the effects that mergers have on the MBH population in CCGs.  Two main factors contribute to their evolution. Firstly, CCGs experience many more dry mergers with spheroids than other galaxies. Parabolic dry mergers grow a galaxy's mass, luminosity and radius more than a galaxy velocity dispersion \citep[e.g.,][] {CvA2001,Nipoti2003,CLV2007,Naab2009,Nipoti2009,Shankar2011,Oser2012,Hilz2012}. If in a given merger $\mbh$ and $M_{\rm bulge}$ increase relatively more than $\sigma$ (see the discussion in Ciotti et al. 2007; Ciotti 2009), a sequence of such mergers will lead to more massive MBHs  at fixed $\sigma$. In Fig.~2 (left panels) this corresponds to moving upward more efficiently than rightward. 

Secondly, the sheer number and mass contribution of MBH-MBH mergers occurring in CCGs galaxies is much higher than in other galaxies (see Table~1). If we assume that correlations between MBHs and hosts are established because of AGN feedback \citep[e.g.,][]{Silk1998,Fabian1999,DiMatteo2005,Hopkins2009}, then if MBH mergers contribute to the MBH growth {\it after} the bulk of quasar/AGN activity has ceased (Fig.~1), the MBH mass increase brought by these mergers will then push the MBH upwards and out from the $\mbh-\sigma$ correlation established through feedback.  One important caveat, however, is whether MBH binaries can merge efficiently in gas-poor environments, because of the so-called `final parsec problem' \citep[e.g.,][]{BBR1980}, although various effects, such as triaxiality and rotation, as well as the presence of massive perturbers, may increase the orbital decay rate, see \cite{Colpi2011} for a recent review. The presence of multiple MBHs may also occur \citep{2002MNRAS.336L..61H,2003ApJ...593..661V,2012MNRAS.422.1306K}

Our models are at variance with other models that study the impact of MBH mergers on the establishment of correlations \citep[e.g.,][]{Jahnke2011} as we do not assume that MBHs populate all galaxies. In fact, the presence or absence of a central MBH leads to different evolutionary paths in the $\mbh-\sigma$ and $\mbh-M_{\rm bulge}$ relations. We can consider two extreme cases. Let us assume that all  galaxies host a MBH. Then at each merger $\mstar$ and $\mbh$ increase as the sum of those in the two galaxies, barring for the effects of stellar escapers \citep{Nipoti2003,Hilz2012} and non-linear addition of MBH masses \citep{CvA2001,CLV2007}. However,  Eq.~3 shows that $\sigma$ would stay the same or slightly decrease. This corresponds to a vertical upward movement in the $\mbh-\sigma$ plot, eventually leading to MBHs that are over-massive for their $\sigma$, but are not outliers in the $\mbh-M_{\rm bulge}$ correlation. The other extreme case assumes that only the main galaxy hosts a MBH. Then at each merger $\mstar$ increases, while $\mbh$ and $\sigma$ do not. Eventually the MBH in the galaxy that results from the merger sequence will be under-massive for its bulge mass, but it will not be an outlier in the $\mbh-\sigma$ relation.

Broadly speaking, the models of MBH evolution that we adopt for this paper (Volonteri et al. 2012) predict that the most massive MBHs, except for those hosted in CCGs, are those that are best correlated with their hosts \citep{VN09} if their build-up is driven by a combination of accretion and mergers that includes both gas-rich and gas-poor galaxies, and dispersion should increase at low MBH/galaxy masses \citep[see Fig.~2, top panel in][]{Volonteri2012}, where the MBH mass, even at $z=0$ traces the properties of the MBH formation mechanism \citep{Svanwas2010}. It may well be that if different processes shape the MBH mass at different galaxy masses (MBH formation at the lowest masses, AGN feedback at intermediate masses, MBH and dry mergers at the highest masses), there is not a unique link that straddles throughout the whole range.

\acknowledgements 
We are warmly grateful to C. Nipoti and K. G\"ultekin for insightful comments.  MV acknowledges funding support from NASA, through Award Number ATP NNX10AC84G; from SAO, through Award Number TM1-12007X, from NSF, through Award Number AST 1107675, and from a Marie Curie Career Integration grant (PCIG10-GA-2011-303609). L.C. aknowledges financial support from PRIN MIUR 2010-2011, project ``The Chemical and Dynamical Evolution of the Milky Way and Local Group Galaxies'', prot. 2010LY5N2T.


\end{document}